\documentstyle{article}
\title{\bf Vertex Models on Feynman Diagrams
}
\author{ {\it D.A. Johnston}\\
         Dept. of Mathematics\\
         Heriot-Watt University\\
         Riccarton\\
         Edinburgh, EH14 4AS, Scotland\\ \\
and\\ \\
         {\it P. Plech\' a\v{c}}\\
         Mathematical Institute\\
         24-29 St Giles'\\
         Oxford\\
         OX1 3LB}
\date {29th August 1997}         
\begin{document}
  \maketitle
                      {\Large
                      \begin{abstract}
%
The statistical mechanics of spin models, such as the
Ising or Potts models, 
on generic random graphs can be formulated 
economically by considering  the 
$N \rightarrow 1$ limit of $N \times N$ Hermitian
matrix models. In this paper we consider
the $N \rightarrow 1$  limit in {\it complex}
matrix models, which describes vertex models
of different sorts living on random graphs.
From the graph theoretic perspective
one is using matrix model and field theory
inspired methods to count various classes
of directed graphs. 

We also make some remarks
on vertex models on planar
random graphs (the $N \to \infty$ limit) where the resulting
matrix models are not generally soluble using currently known
methods. Nonetheless, some particular cases may be mapped
onto known models and hence solved.
%
                        \end{abstract} }
%
  \thispagestyle{empty}
%
%
  \newpage
%
                  \pagenumbering{arabic}

\section{Introduction}

In recent papers we have discussed various analytical and numerical
aspects of an economical way of describing spin
models living on ``thin'' (i.e. generic, non-planar) random graphs \cite{1,1a}.
In particular, one finds mean-field like behaviour for the spin models
due to the locally-tree-like structure of the graphs.
This approach is based on the simple observation \cite{2} that the thin graphs
appear as the Feynman diagrams in the 
perturbative expansion of matrix models when the $N \to 1$ (scalar) limit is taken.
An alternative  $N \to \infty$ limit is
perhaps more familiar for matrix models in the context of two-dimensional gravity,
where the resulting ``fat'' graphs
are of interest because of their relation to surfaces
and string worldsheets. 
This relation is lost in the thin graph case \footnote{The surfaces
that might be associated with the graphs are ``infinite genus'',
with a maximal handle density.} but
the models still merit attention, both from the point of view of
investigating various decorated random graphs and as a 
way of accessing mean field behaviour without infinite range interactions
or the boundary problems of genuine trees like the Bethe lattice. 

The thin graph models are amenable
to a saddle point solution, using large order/instanton
methods from path integrals \cite{3}
for the ordinary
integrals which appear in this case. 
Such saddle point
methods in graph theory been independently derived
from a probabilistic  
viewpoint by Whittle in \cite{4}.
The topic of the current paper has also been presaged by
the same author
in \cite{5}, where the main concern was understanding the statistics
of random directed graphs 
using complex integrals rather than the 
real integrals of \cite{1,1a,4}. We take a rather different tack
here, where our aim is to 
investigate vertex models on random graphs in their own right,
so the focus is on the matter living on the graphs rather than
the graphs themselves.

In diagrammatic terms the effect
of going from a Hermitian to a complex matrix model
is to place arrows on the edges of the graphs that
appear in the perturbative expansion, giving 
a directed graph. 
We take $\phi^{\dagger}$ to be
the ``head''  of the arrow by convention.
This is true whatever
the value of $N$ so it still holds, in particular,
for the $N \to 1$ 
limit. We can pick out the $2n$th order
in the perturbative expansion by carrying out
a contour integral in the coupling constant
in an exactly analogous manner to the 
path integral calculations in \cite{3}.
The general form of the partition function we
are interested in for a vertex model
on an ensemble of thin random graphs with $2n$ vertices is thus
\begin{equation}
Z_n(\{\Lambda\}) \times N_n = {1 \over 2 \pi i} \oint { d \lambda \over
\lambda^{2n + 1}} \int {d \phi^{\dagger} d \phi \over 2 \pi }
\exp \left[- \left( \frac{1}{2} \phi^{\dagger} \phi  + P (\phi^{\dagger}, 
\; \phi, \; \lambda, \; \{ \Lambda \} ) \right) \right],
\label{e1}
\end{equation}
where $\phi$ is complex, $\phi^{\dagger}$ is its conjugate, $P(\phi^{\dagger}, \; \phi, \; \lambda, \; \{ \Lambda \} )$
is the potential that specifies the particular vertex model, $\lambda$ is the overall vertex coupling that will 
invariably be
scaled out in the ensuing discussion and $\{ \Lambda \}$ is the set of remaining couplings that 
can be tuned to search for critical behaviour.
The prefactor $N_n$ counts the number of undecorated graphs in the class of graphs of interest and will
generically grow factorially with the number of vertices. For $\phi^3$ (3-regular) random graphs, for instance,
it is given by
\begin{equation}
N_n = \left( {1 \over 6} \right)^{2n} { ( 6 n - 1 ) !! \over ( 2 n ) !!
}.
\end{equation}

It is well known that complex matrix models with
potentials of the form $P( (\phi^{\dagger} \phi), \; \lambda, \; \{ \Lambda \})$
generate ``chequered''
surfaces in the $N \to \infty$ limit, where all the loops have an even number of edges
\cite{6}. In this case 
the critical behaviour turns out to be identical
to that of a Hermitian model, as restricting oneself to combinations
of $\phi^{\dagger} \phi$ effectively gives a Hermitian model in terms
of $| \phi | $. Something similar occurs in the thin $N \to 1$ limit,
as was noted already in \cite{5}.  When one considers models
where the in-degree (number of arrows in) and out-degree (number of arrows out)
at each vertex is the same, as would be the case for potentials
of the form $P( (\phi^{\dagger} \phi), \; \lambda, \; \{ \Lambda \})$
the saddle point equations are singular in the original variables due to
the radial degree of freedom
\footnote{An identical situation occurs
in the path integral for coupled anharmonic
oscillators, where the 
large orders saddle point solution has a divergence
at the radially symmetric point
\cite{7}.}. Switching to polar co-ordinates 
eliminates the singularity and one recovers the critical behaviour of
the equivalent real integrals.

The natural symmetry to preserve in the vertex models
is arrow reversal, or complex conjugation.
In the integral for the partition function in equ.(\ref{e1})
this corresponds to demanding a real integrand.
We now note that we are not obliged to restrict ourselves
to potentials composed only of combinations of $\phi^{\dagger} \phi$ 
to maintain this symmetry, which allows us to look at a 
more general class of potentials and observe a much richer
critical behaviour than that 
seen for $P( (\phi^{\dagger} \phi), \; \lambda, \; \{ \Lambda \})$.
We shall see in the sequel
that the resulting models play a similar role to the eight-vertex
model on a square lattice in that they contain many other soluble
models as special cases.

In the next two sections of the paper we discuss models on thin $\phi^3$ 
and $\phi^4$ graphs separately.
In all cases we shall convert our models back to a
real notation by making the substitution
$\phi = x + i y, \; \phi^{\dagger} = x - i y$ ($x,y$ real), as well
as pre-emptively scaling out the overall vertex coupling $\lambda$ in 
the actions we discuss. As it is the
critical behaviour of the matter that is of interest, 
which is determined by the saddle point equations for
the ``fields'' ($\phi^{\dagger}, \; \phi$), we shall
concentrate exclusively on them,
omitting the trivial saddle point in $\lambda$.
We then move on to discuss vertex models on {\it planar}
graphs (i.e. the $N \to \infty$ limit), where the resulting matrix models 
are not soluble in general, highlighting some cases
where solutions are available by mapping onto previously solved models.

\section{Vertex Models on (Thin) $\phi^3$ Graphs}

On $\phi^3$ graphs 
we can have vertices with 3 inward or 3 outward pointing arrows,
2 arrows in and 1 out, or 2 arrows out and one in
as shown in Fig.1.
These correspond to terms of the form
$(\phi^{\dagger})^3$, $\phi^3$, $(\phi^{\dagger})^2 \phi$ and $\phi^{\dagger}
(\phi)^2$ respectively in $P(\phi^{\dagger}, \;  \phi, \; \lambda, \; \{ \Lambda \})$. 
We can construct four possible linear combinations
of these that are invariant under the complex conjugation symmetry,
namely
\begin{eqnarray}
&{}& \left[(\phi^{\dagger})^3 + \phi^3 \right], \; \; \; 
i \left[(\phi^3 - (\phi^{\dagger})^3 \right] \nonumber \\
&{}& \left[(\phi^{\dagger})^2 \phi + \phi^2 (\phi^{\dagger}) \right] , \; \; \;
i \left[(\phi^{\dagger})^2 \phi - \phi^2 (\phi^{\dagger}) \right] 
\end{eqnarray}
so the most general invariant action $S$
($= \frac{1}{2} \phi^{\dagger} \phi + P(\phi^{\dagger}, \;  \phi, \; \lambda = 1, \; \{ \Lambda \})$) one can write
down is
\begin{eqnarray}
S = \frac{1}{2} \left\{ \phi^{\dagger} \phi - { \alpha \over 3} \left[(\phi^{\dagger})^3 + \phi^3 \right]
- i { \beta \over 3} \left[(\phi^3 - (\phi^{\dagger})^3 \right] 
-  \gamma \left[(\phi^{\dagger})^2 \phi + \phi^2 (\phi^{\dagger}) \right]
- i \delta \left[(\phi^{\dagger})^2 \phi - \phi^2 (\phi^{\dagger}) \right] \right\}
\end{eqnarray}
or, grouping together the similar vertices
\begin{eqnarray}
S = \frac{1}{2} \left\{ \phi^{\dagger} \phi - { (\alpha - i \beta) \over 3}  (\phi^{\dagger})^3
- { (\alpha + i \beta) \over 3} \phi^3 - (\gamma + i \delta) \left[ (\phi^{\dagger})^2 \phi \right]
- (\gamma- i \delta) \left[ \phi^2 (\phi^{\dagger}) \right] \right\} .
\label{ecomp}
\end{eqnarray}
We thus see it is natural to have the $(\phi^{\dagger})^3, \; \phi^3$
and $(\phi^{\dagger})^2 \phi, \; \phi^2 (\phi^{\dagger})$ vertices
appearing with complex (but conjugate) weights.
We can cast this action in real form by writing
$\phi = x + i y, \; \phi^{\dagger} = x - i y$, with $x$, $y$ real
which gives
\begin{eqnarray}
S = \frac{1}{2} ( x^2 + y^2) -  { (\alpha + 3 \gamma) \over 3} x^3 -  { (\beta + 3 \delta) \over 3}  y^3
- ( \gamma -  \alpha ) x y^2 -  ( \delta -  \beta ) x^2 y.
\label{ereal}
\end{eqnarray}
This transformation is rather reminiscent of the line decomposition
that is used for vertex models on regular lattices \cite{8},
with $x$ being black and $y$ white lines,
but we do not have line conservation at each vertex
for the $\phi^3$ model.
The saddle point equations $\partial S / \partial x, \; \partial S / \partial y = 0$ 
for this action are explicitly soluble for general values of
the couplings, but the solutions are not terribly illuminating 
for the purposes of extracting different critical
behaviours when written down baldly.

We can get a little more insight by reducing various subsets of the coupling constant space
to already known models. One obvious reduction is to set $\beta = \delta = 0$
which removes the terms antisymmetric in $(\phi^{\dagger}, \phi)$ in the action. This leaves one  
with 
\begin{equation}
S = \frac{1}{2} ( x^2 + y^2) -   {( \alpha + 3 \gamma  ) \over 3}  x^3  - ( \gamma -  \alpha ) x y^2
\label{ising1}
\end{equation}
which displays mean-field Ising critical behaviour
on a suitable coupling constant locus. 

This can be seen
by considering the standard action for the Ising model on $\phi^3$ graphs
\begin{equation}
S = \frac{1}{2} ( x^2 + y^2) - c  x y   - \frac{1}{3} ( x^3 + y^3)
\label{ephi3}
\end{equation}
(where $c = \exp ( - 2 \beta)$)
whose Feynman diagram expansion represents Ising spins
with Hamiltonian
\begin{equation}
H = \beta \sum_{ij} \sigma_i \sigma_j
\end{equation} living on the graphs,
where the spin interactions are along nearest neighbour $<ij>$ edges.
If we 
carry out the linear transformation $x \to \frac{1}{\sqrt{2}} ( x + y), \; y
\to \frac{1}{\sqrt{2}} ( x - y )$ and some judicious rescalings we obtain
\begin{equation}
S = \frac{1}{2} ( x^2 + y^2) - \frac{1}{3} \left[ x^3 + 3 c^*  x y^2 \right],
\label{ising2}
\end{equation}
where $c^* = (1 - c ) / ( 1 + c )$,
which is identical (up to a rescaling) with equ.(\ref{ising1}), provided we identify
$c^*$ with $( \gamma -  \alpha ) / (\alpha + 3 \gamma)$.

The Ising model on $\phi^3$ graphs described by the action in equ.(\ref{ephi3})
displays a mean field transition at $c=1/3$, so the model in equ.(\ref{ising2})
would be expected to display a transition at the ``dual'' value of $c^* = 1/2$.
A direct solution of the saddle point equations for equ.(\ref{ising2})
\begin{eqnarray}
x &=&  1, \; \;   \; y =  0 \; \; \; (Low \; T^*) \nonumber \\
x &=& { 1 \over 2 c^* }, \; \;   \; y =  \sqrt{2 c^* - 1  \over 4 (c^*)^3} \; \; \; (High \; T^*),
\end{eqnarray}
where the $T^*$ is the temperature associated with
the dual coupling, $c^* = \exp ( - 2 \beta^*)$, reveals that this is indeed the case. 
It is worth remarking that the action of equ.(\ref{ising1}, \ref{ising2}) 
on planar graphs (i.e. in the $N \to \infty$ limit)
corresponds to 
an Ising model with spins situated at the vertices of a random
triangulation dual to the original $\phi^3$ graph \cite{9}.
The notion of duality is not defined for the non-planar
graphs of the $N \to 1$ limit but one could think
of action representing spins residing on the vertices of the 
triangles in the cactus
(or Husimi tree) diagram in Fig.2
interacting with their nearest neighbours.
This cactus graph is the medial graph of the underlying
$\phi^3$ graph which is also indicated
in the figure. 
Unlike standard cacti, the branches of triangles eventually
close to form large loops, since the underlying $\phi^3$
graphs are closed.
For an alternative physical interpretation one could
consider the $\phi^3$ graphs themselves and 
think of the model as a gas of $y$ loops
in an $x$ ``sea'', which is the $O(1)$ loop gas
representation of the Ising model.

Viewed as Ising spins on a cactus graph the action of
equ.(\ref{ising1}, \ref{ising2})
restricts triangles to have all spins up ($x^3$
vertices) or
two spins down and one up ($x y^2$ vertices). This
restriction does not appear in the planar case
as one can then think of $x$ representing a ``same-spin''
edge and $y$ a ``different-spin'' edge, which is
sufficient to cover all configurations
when the triangles are pasted together along their edges.
The difference on the thin graphs is
that the triangles are to be thought of as being
joined at their vertices. 
Lifting the constraint is equivalent
to allowing non-zero $\delta$ and $\beta$ in the original vertex
model and we obtain actions that describe
multi-spin interaction  models on cacti.
The spin Hamiltonian for such a model is given by
\begin{equation}
H = \beta ( J_3 \sum_{\Delta} \sigma_i \sigma_j \sigma_k + h \sum_i \sigma_i)
\label{multi1}
\end{equation}
where the three spin sum is over triangles $\Delta$.
With all four possible spin configurations on the triangles
of the Husimi tree now allowed, namely
$(+ \; + \; + ), \; (-\; -\; -), \; (+ \; + \; -), \; (+ \; - \; -)$,
the action can be written as
\begin{equation}
S = {\mu \over 2 }  x^2 + {1 \over 2} y^2 -  \mu^2  x^2 y
    - \zeta \mu  y^2 x  - { \zeta \mu^3 \over 3} x^3 - {1 \over 3} y^3,
\label{multia1}
\end{equation}
where $\mu = \exp ( 2 \beta h)$, $\zeta = \exp ( 2 \beta J_3)$.
This is clearly equivalent to 
equ.(\ref{ereal}) once one rescales $x \to x  / \sqrt{\mu}$
and makes the identifications
\begin{eqnarray}
\beta + 3 \delta  &=& 1 \nonumber \\
\alpha + 3 \gamma &=& \zeta \mu^{3/2} \nonumber \\
\gamma - \alpha &=& \zeta \mu^{1/2} \nonumber \\
\delta - \beta &=& \mu
\end{eqnarray}
which gives two conditions on our original four vertex model couplings
to match the number of physical parameters.
Note that it is natural to include the external field $h$ from the start
in such models as the phase transition generically appears
in non-zero field for such multi-site interaction models \cite{10}.

The Hamiltonian of equ.(\ref{multi1}) may be extended to include nearest neighbour interactions
as well
\begin{equation}
H = \beta ( J_3 \sum_{\Delta} \sigma_i \sigma_j \sigma_k + J_2 \sum_{ij} \sigma_i \sigma_j + h \sum_i \sigma_i).
\label{multi2}
\end{equation}
and a similar reasoning to the pure 3-spin interaction case above leads to the action
\begin{equation}
S = \nu \left( {\mu \over 2 }  x^2 + {1 \over 2} y^2 -  \mu^2
 x^2 y
    - \zeta \mu  y^2 x - { \zeta \nu^2 \mu^3 \over 3} x^3 - {\nu^2
 \over 3} y^3 \right),
\label{multia2}
\end{equation}
where the additional parameter $\nu=\exp ( 2 \beta J_2)$. Once again
a rescaling $x \to x  / \sqrt{\mu}$ and absorbing the 
overall factor of $\nu$ in the vertex coupling gives an action of the form
equ.(\ref{ereal}). We can also see that sending $J_3 \to \infty$ ($\zeta \to \infty$)
picks out the two sorts of vertex/triangle allowed in the Ising model discussed above.
From the loop gas point of view both equ.(\ref{multia1},\ref{multia2})
represent a gas of $y$ loops and lines in an $x$ sea, which can now interact via $y^3$
terms and break via $x^2 y$ terms. Alternatively the roles of $y$ (loops/lines)
and $x$ (sea) may be exchanged
as their vertices appear in a similar form in the action.

It is also possible to bring Potts models under the umbrella of the vertex model framework
in one of two possible ways.
The original $q$-state Potts model action contains $q$ (real) variables $\phi_i$
\begin{equation}
S = { 1 \over 2 } \sum_{i=1}^{q} \phi_i^2 - c \sum_{i<j} \phi_i \phi_j -{1 \over 3} \sum_{i=1}^q \phi_i^3,
\label{qstate}
\end{equation}
where $c$ is now $1/ (\exp( 2 \beta ) + q - 2)$,
and represents the following Hamiltonian on $\phi^3$ graphs
\begin{equation}
H =   \beta \sum_{<ij>} ( \delta_{\sigma_i, \sigma_j} -1),
\end{equation}
where the spins $\sigma_i$ take on $q$ values and $\delta$
is a Kronecker delta.
It is possible to write an ``effective'' action coming from
imposing the observed symmetry breaking pattern 
$\phi_{1 \ldots q-1} = x, \phi_q = y$ from the saddle point equation solutions
on the original $q$ variable action \cite{1}. This effective action
\begin{eqnarray}
S = {1 \over 2} ( q - 1) \left[ 1 - c ( q - 2) \right] x^2  - { 1 \over 3} ( q -1) x^3
+ {1 \over 2} y^2  - {1 \over 3} y^3 - c ( q -1 ) x y
\label{app1}
\end{eqnarray}
faithfully
reproduces the correct saddle point solutions.
One can carry now out some rescalings on equ.(\ref{app1})
along with the ubiquitous linear transformation
$x \to (x - y ) / \sqrt{2}$, $ y \to ( x + y ) / \sqrt{2}$
to obtain
\begin{eqnarray}
S = \frac{1}{2} ( x^2 + y^2) -  { 1  \over 3} \left[ x^3
+  3 \Omega \Delta  x^2 y +   3 \Omega^2 x y^2 + \Omega^3 \Delta y^3 \right]
\label{ereal2}
\end{eqnarray}
where
\begin{eqnarray}
\Omega &=& \sqrt{ { \sqrt{\chi} - \kappa \over \sqrt{\chi} + \kappa}} \nonumber \\
\chi &=& (q-1) ( 1 - c ( q - 2)) \; , \; \; \; \kappa = c ( q-1 ) \nonumber \\
\Delta &=& { ( q - 1) - \chi^{3/2} \over ( q - 1) + \chi^{3/2}}
\end{eqnarray}
which is yet another variant of the action in equ.(\ref{ereal}) for a particular
choice of couplings. It is noteworthy that the 
first order transition point for the $q$-state Potts model that was
derived in \cite{1}
\begin{equation}
c = { 1 - (q-1)^{-1/3} \over  q - 2}
\end{equation}
is the point at which $\Delta=0$, reducing equ.(\ref{ereal2}) to 
an  
Ising-like action .

A second possibility is to include two further quadratic terms in the allowed action
\begin{eqnarray}
\frac12 [ \phi^2 + ( \phi^{\dagger} )^2 ]  &=& x^2 - y^2 \nonumber \\
\frac{i}{2} [ \phi^2 - ( \phi^{\dagger} )^2 ]  &=& - 2 x y .
\end{eqnarray}
which are best thought of as addition bivalent vertices
from the point of view of maintaining a vertex model interpretation.
These allow two arrow heads or tails to meet.
With these additional vertices the action becomes
\begin{eqnarray}
S &=& \frac{1}{2} \left\{ \mu \phi^{\dagger} \phi - ( \nu + i \xi)   \phi^2 
- (\nu - i \xi)   ( \phi^{\dagger} )^2  \right. \nonumber \\
&-&  \left. { (\alpha - i \beta) \over 3}  (\phi^{\dagger})^3
- { (\alpha + i \beta) \over 3} \phi^3 - (\gamma + i \delta) \left[ (\phi^{\dagger})^2 \phi \right]
- (\gamma- i \delta) \left[ \phi^2 (\phi^{\dagger}) \right] \right\} .
\label{epotts}
\end{eqnarray}
and choosing 
$\alpha = \gamma = ( q - 1)/4 $, $\beta= \delta = 1/ 4$, $\xi = -c ( q -1 ) / 2$,
$\mu = [ 1 + (q - 1) ( 1 - c ( q - 2)\; )]/2$, $\nu = [ 1 - (q - 1) ( 1 - c ( q - 2) \; )]/4$
gives an action which is recognisable as that for a $q$-state Potts model 
in equ.(\ref{app1}) when written 
in real form.

To summarize the results of this section - the action of equ.(\ref{ecomp})
for a  vertex model on thin $\phi^3$ graphs has been shown to contain as special
cases Ising and multi-spin interaction models, results which are easiest 
to see once the model has been rewritten in the real form of
equ.(\ref{ereal}). 
Written is this form the vertex models can be re-interpreted as representing various 
systems,
including spins on cacti or loop/line gases on the original $\phi^3$ graphs.
Potts models may also be incorporated into the general framework,
either by reconstructing the effective Potts model
action by rescaling and linear transformations from equ.(\ref{ereal}),
or by introducing further quadratic vertices as in equ.(\ref{epotts}).

\section{Vertex Models on (Thin) $\phi^4$ Graphs}

We deal with the $\phi^4$ vertex models rather more briefly, as the essential
features are similar to the $\phi^3$ models in the previous section.
For $\phi^4$ graphs one has the option
of being able to preserve the in-degree and out-degree 
of each sort of vertex, but as we have noted this is equivalent
to considering potentials composed of products of $\phi^{\dagger} \phi$
and exploring a rather restricted class of critical behaviours.
In a similar manner to the $\phi^3$ graphs,
we therefore
consider all the conjugation symmetric terms
\begin{eqnarray}
&{}& \left[(\phi^{\dagger})^4 + \phi^4 \right], \; \; \;
i \left[(\phi^4 - (\phi^{\dagger})^4 \right] \nonumber \\
&{}& \left[(\phi^{\dagger})^2 \phi^2 \right] , \; \; \;
\nonumber \\
&{}& \left[(\phi^{\dagger})^3 \phi + \phi^3 (\phi^{\dagger}) \right] , \; \; \;
i \left[(\phi^{\dagger})^3 \phi - \phi^3 (\phi^{\dagger}) \right].
\end{eqnarray}
The most general action containing all these terms is
\begin{eqnarray}
S & = & \frac12 \left\{ \phi^{\dagger} \phi - {\alpha \over 4} \left[ (\phi^{\dagger})^4 + \phi^4 \right]
- i {\beta \over 4} \left[ \phi^4 - (\phi^{\dagger})^4  \right]
 - \gamma (\phi^{\dagger})^2 \phi^2  \right. \nonumber \\
 & - &  \left. \sigma \left[ (\phi^{\dagger})^3 \phi + \phi^3 (\phi^{\dagger}) \right] - 
     i \rho \left[ (\phi^{\dagger})^3 \phi - \phi^3 (\phi^{\dagger}) \right] \right\}
\label{ecomp4}
\end{eqnarray}
or 
\begin{eqnarray}
S  =  \frac12 \left\{ \phi^{\dagger} \phi - {( \alpha  + i \beta ) \over 4}  \phi^4
-  {( \alpha - i \beta )   \over 4}  (\phi^{\dagger})^4  
 - \gamma (\phi^{\dagger})^2 \phi^2  
  -   ( \sigma  + i \rho )   [ (\phi^{\dagger})^3 \phi]  -
     ( \sigma   - i \rho )    [ \phi^3 (\phi^{\dagger}) ] \right\}.
\label{ecomp4a}
\end{eqnarray}
As for the $\phi^3$ examples, this is  best digested in bite-size chunks.
Switching to real notation
\begin{eqnarray}
S  &=&  \frac12 ( x^2 + y^2 ) - { \left( \alpha  + 2 \gamma + 4 \sigma \right) \over 4}  x^4
- { \left( \alpha  + 2 \gamma - 4 \sigma \right) \over 4}   y^ 4- 
 -   { ( 3 \alpha + 2 \gamma ) \over 2 }  x^2  y^2 \nonumber \\
& -&  ( 2 \rho +  \beta)  x^3  y 
-  ( 2 \rho -  \beta )  x  y^3
\label{ereal4}
\end{eqnarray}
we again find an action that has Ising and multi-spin interaction critical behaviour
for various reduced subsets of its coupling constants. Setting 
the asymmetric terms $\rho$
and $\beta$ to zero again gives mean-field Ising critical behaviour, as can be seen
in an exactly analogous fashion to the $\phi^3$ graphs by carrying
out the linear transformation
$x \to \frac{1}{\sqrt{2}} ( x + y), \; x \to
\frac{1}{\sqrt{2}} ( x - y )$ on the action for an Ising model living
on $\phi^4$ graphs
\begin{equation}
S = \frac{1}{2} ( x^2 + y^2) - c x y   - \frac{1}{4} ( x^4 + y^4)
\label{ephi4}
\end{equation}
to get (with suitable scalings) 
\begin{equation}
S = \frac{1}{2} ( x^2 + y^2 ) - \frac14 x^4 - \frac32 c^* x^2 y^2 - \frac14 (c^*)^2 y^4.
\label{edual4}
\end{equation}
The original $\phi^4$ action displays a mean-field Ising phase transition at $c = 1/2$, so one would expect
the transformed action to have a transition at $c^* = 1/3$, which a direct solution
of the saddle point equations for equ.(\ref{edual4}) 
\begin{eqnarray}
x &=&  1, \; \;   \; y =  0 \; \; \; (Low T^*) \nonumber \\
x &=& { 1 \over 2  } \sqrt{{3 - c^* \over 2 c^*}}, \; \;   \; y =  
 {\sqrt{3 c^* - 1  \over 8 (c^*)^{2}}} \; \; \; (High T^*)
\end{eqnarray}
confirms, just as for the $\phi^3$ Ising model. 

Once again the Ising-like actions are seen to reside in the 
symmetric subspace of couplings. Allowing
non-zero $\rho$ and $\beta$ in equs.(\ref{ecomp4},\ref{ereal4}) gives 
actions that may be interpreted as representing either multi-spin
interactions
on cacti composed of squares, such as that shown in Fig.3,
or loop/line gases on the original $\phi^4$ graphs.
The spin model Hamiltonians on cacti graphs represented
by such actions are of the form
\begin{equation}
H = \beta ( J_4 \sum_{[i,j,k,l]} \sigma_i \sigma_j \sigma_k \sigma_l + J_2 \sum_{ij} \sigma_i \sigma_j + h \sum_i \sigma_i),
\label{multi4}
\end{equation}
where the sum $[i,j,k,l]$ is round the squares of the cactus,
with similar relations between the spin couplings $(J_4, J_2, h, \beta)$ and the coefficients
in the action equ.(\ref{ereal4}) as in the $\phi^3$ case.

\section{Vertex Models on Planar (Fat) Graphs}

Writing down vertex model actions for the $N \to \infty$ limit
is a simple matter of putting a $Tr$ in front of the previously
discussed actions such as equ.(\ref{ecomp}) and interpreting
the variables as $N \times N$ matrices rather than scalars.
In the case of higher than $\phi^3$ potentials
the non-commutativity of the matrix variables will also
allow the introduction of new terms compared with the
thin graphs as, for example, terms such as
$[\Phi^{\dagger} \Phi ]^2$ and $[\Phi^{\dagger}]^2 \Phi^2$,
where the $\Phi, \Phi^{\dagger}$ are now $N \times N$ matrices,
may be distinguished. This difference from the thin graph case
arises because a consistent cyclic order of legs
round a vertex may be defined in the planar limit which makes 
such distinctions meaningful, as can be seen in Fig.4.
At the level of the partition function an explicit contour integral
to pick out the $2n$th order is no longer necessary as the number of planar 
graphs increases only exponentially with the order, so the
vertex coupling (``cosmological constant'') may be tuned to get 
a diverging number of vertices. This also means that the factorial
prefactor $N_n$ need no longer be peeled off.
Writing down such partition functions is one thing, solving them is another
as one is no longer dealing with saddle points
in simple scalar integrals.
In general the result is an insoluble, or at least unsolved, matrix model.
Nonetheless, the planar graph vertex models can be mapped onto
other, solved, models in some particular cases
which we discuss below.

The non-commutativity of the matrix variables
in the planar fat graph limit $N \rightarrow \infty$
plays no role on $\phi^3$ graphs, as the 
cyclic symmetry of the trace ensures that possible
potential terms are identical to the thin graph case.
We have, writing $\Phi = X + i Y$ for the $N \times N$
complex matrix $\Phi$ and $N \times N$ Hermitian matrices
$X,Y$ 
\begin{eqnarray}
Tr \left[(\Phi^{\dagger})^2 \Phi + \Phi^2 (\Phi^{\dagger}) \right]
&=& 2 \; Tr \left[ X^3 + X Y^2 \right] \nonumber \\
i \; Tr \left[(\Phi^{\dagger})^2 \Phi - \Phi^2 (\Phi^{\dagger}) \right]
&=& 2 \; Tr \left[ Y^3 + Y X^2 \right] \nonumber \\
Tr \left[ (\Phi^{\dagger})^3 + \Phi^3 \right] &=&
2 \; Tr \left[ X^3 - 3 X Y^2 \right] \nonumber \\
i \; Tr \left[ \Phi^3  - (\Phi^{\dagger})^3 \right] &=&
2 \; Tr \left[ Y^3 - 3 Y X^2 \right],
\end{eqnarray}
just as  for thin graphs. The fat graph version of the
action in equs.(\ref{ecomp},\ref{ereal}) therefore
differs only in the presence of a trace
\begin{eqnarray}
S = Tr \left\{ \frac{1}{2} ( X^2 + Y^2) -  { (\alpha + 3 \gamma) \over 3} X^3 -  { (\beta + 3 \delta) \over 3}  Y^3
- ( \gamma -  \alpha ) X Y^2 -  ( \delta -  \beta )X^2 Y \right\}.
\label{erealfat}
\end{eqnarray}
Equ.(\ref{erealfat}) is known to represent the Ising model on dynamical triangulations (i.e. the dual
of planar $\phi^3$ graphs) with the spins situated at the triangle vertices, when $\beta=\delta=0$ 
and $( \gamma -  \alpha ) / (\alpha + 3 \gamma) = c^*$ \cite{9}.
Its critical behaviour is therefore that of the Ising model coupled to 2D gravity, giving the KPZ critical
exponents \cite{11} rather than the flat 2D lattice Onsager exponents.
We can say little about the more general case, as the solution of the resulting matrix model   
is not known.

For fat $\phi^4$ graphs the non-commutativity {\it does}
play a role as there are now more quartic invariants
than in the thin case. 
Writing $\Phi = X + iY$  again,
we have the following possibilities
\begin{eqnarray}
Tr \left[(\Phi^{\dagger})^2 \Phi^2  \right]
&=&  Tr \left[ X^4 +  Y^4  + 2 X Y X Y\right] \nonumber \\
Tr \left[ \left( (\Phi^{\dagger}) \Phi \right)^2 \right] &=&
Tr \left[ X^4 + Y^4  + 4 X^2 Y^2 - 2 XYXY \right]
\nonumber \\
Tr \left[(\Phi^{\dagger})^4 + \Phi^4 \right]
&=& 2 \; Tr \left[ X^4 + Y^4 - 4 X^2 Y^2 - 2 XYXY \right]
\nonumber \\
i \; Tr \left[\Phi^4  - (\Phi^{\dagger})^4 \right]
&=& 8 \; Tr \left[ X Y^3 -  X^3 Y\right]
\nonumber \\
Tr \left[(\Phi^{\dagger})^3 \Phi + \Phi^3 (\Phi^{\dagger})
\right] &=&
2 \; Tr \left[ X^4 - Y^4 \right] \nonumber \\
i \; Tr \left[(\Phi^{\dagger})^3 \Phi - \Phi^3 (\Phi^{\dagger})
\right] &=&
 4 \; Tr \left[ X^3 Y + Y^3 X \right].
\label{ePhi4dict}
\end{eqnarray}
Dropping the terms 
containing an asymmetric combination of $\Phi, \Phi^{\dagger}$
it is possible
to construct an action that is equivalent to the 
Ising action for spins
on the vertices of a random quadrangulation
(i.e. the duals of $\phi^4$ graphs) 
by choosing the appropriate combination of
the remaining terms 
\begin{equation}
S =  Tr \left\{ {1 \over 2} X^2  + { 1 \over 2} Y^2 
- {1 \over 4} X^4  -   {c^* \over 2} ( X Y X Y + 2 X^2 Y^2 )  - {(c^*)^2 \over 4} Y^4 \right\},
\label{eplanar4}
\end{equation}
where $c^*$ is again the dual to to the Ising coupling $c =
\exp( - 2 \beta)$, $c^* = (1-c)/(1+c)$.
That equ.(\ref{eplanar4}) {\it does} represent an Ising model
can be shown by retracing the steps that lead from equ.(\ref{ephi4})
to equ.(\ref{edual4}) in the thin graph case, or by
deriving the weights in equ.(\ref{eplanar4})
from direct consideration of the Ising model
on the random quadrangulation \cite{9}.
The most general $\phi^4$ vertex model containing a linear combination of the terms
in equ.(\ref{ePhi4dict}) remains unsolved, but some further examples
can be solved, or at least reduced to eigenvalue integrals and solved in principle.

One of these was noted some years ago in \cite{12}
by Ginsparg. If we consider the action
\begin{equation}
S = Tr \left\{ \frac12 \Phi^{\dagger} \Phi - \alpha
\left[ \left( \Phi^{\dagger} \Phi
\right)^2 + (\Phi^{\dagger})^2 \Phi^2 \right] \right\}
\end{equation}
the resulting Hermitian matrix model action is
\begin{equation}
S = Tr \left\{ \frac12 \left( X^2 + Y^2 \right)
- 2 \alpha \left( X^2 + Y^2 \right)^2 \right\}
\end{equation}
which can be seen to be an $O(2)$/$XY$ 
model by introducing a further
Hermitian matrix $M$ and
rewriting the potential term
to give
\begin{equation}
S = Tr \left\{ \frac12 \left( X^2 + Y^2 + M^2 \right)
- 2 \sqrt{\alpha} ( X^2 + Y^2 ) M \right\}.
\end{equation}
 
Allowing oneself the luxury of two couplings
rather than the one of the $XY$ model also
gives rise to an interesting class of models.
The combination
\begin{equation}
S = Tr \left\{\frac12 \Phi^{\dagger} \Phi - \alpha
(\Phi^{\dagger})^2 \Phi^2 - \beta
\left[ ( \Phi^{\dagger} \Phi )^2 \right] \right\}
\end{equation}
which is equivalent to the real action
\begin{equation}
S = Tr \left\{ \frac12 \left( X^2 + Y^2 \right)
- ( \alpha + \beta ) ( X^4 + Y^4) - 2 ( \alpha - \beta )
X Y X Y - 4 \beta X^2 Y^2 \right\}
\label{eXYXY0}
\end{equation}
can be shown to be a free fermion model \cite{12a}.

Taking another combination
\begin{equation}
S = Tr \left\{\frac12 \Phi^{\dagger} \Phi - \alpha
(\Phi^{\dagger})^2 \Phi^2 - \frac{\beta}{2}
\left[ (\Phi^{\dagger})^4 + \Phi^4 \right] \right\}
\end{equation}
gives
\begin{equation}
S = Tr \left\{ \frac12 \left( X^2 + Y^2 \right)
- ( \alpha + \beta ) ( X^4 + Y^4) - 2 ( \alpha - \beta )
X Y X Y + 4 \beta X^2 Y^2 \right\}
\label{hohum}
\end{equation}
which is similar to the free fermion model apart
from the sign of the $X^2 Y^2$ coupling.
 
If we take $\beta$ negative and $\alpha = - \beta$
in equation.(\ref{hohum})
we get the action
\begin{equation}
S = Tr \left\{\frac12 \left( X^2 + Y^2 \right)
 - 4 | \beta | ( X Y X Y + X^2 Y^2 ) \right\}
\label{eXYXY1}
\end{equation}
considered by Cicuta et.al. \cite{13}
in their discussion of a three-edge colouring problem
for planar graphs.
Although not yet solved exactly the partition function for this
model can be reduced to an eigenvalue integral
and its properties explored
\begin{equation}
Z = \int \prod_i d \lambda_i \prod_{i<j} ( \lambda_i -\lambda_j)^2
\exp \left( -\frac12 \sum_i \lambda_i^2 \right)
\prod_{ij} \left( 1 - 4 | \beta | ( \lambda_i + \lambda_j)^2
\right)^{-\frac12}
\label{ewot}
\end{equation}
where the $\lambda_i$ are the $N$ eigenvalues
of the $N \times N$ matrices.
The scaling of the edge of the eigenvalue
distribution of this
model in the $N \to \infty$ limit
suggest that the critical behaviour is Ising-like,
as the exponents are identical.

The {\it same} eigenvalue expression arises
on taking $\alpha = \beta$ ($\beta$ positive)
in equation.(\ref{hohum}) to give the action
\begin{equation}
S = Tr \left\{ \frac12 \left( X^2 + Y^2 \right)
- 2 \beta (X^2 - Y^2)^2 \right\}
\label{eXYXY2}
\end{equation}
Adopting similar tactics to the $XY$ model and introducing a
further matrix $M$ gives
\begin{equation}
S = Tr \left\{ \frac12 \left( X^2 + Y^2 + M^2 \right)
- 2 \sqrt{\beta} ( X^2 - Y^2 ) M
\right\}
\end{equation}
or
\begin{equation}
S = Tr \left\{ \frac12 M^2 + \frac12 X^2 ( 1 - 2 \sqrt{\beta} M )
+ \frac12 Y^2 ( 1 + 2 \sqrt{\beta} M ) \right\}
\end{equation}
whose partition function reduces to the eigenvalue
integral
\begin{equation}
Z = \int \prod_i d \lambda_i \prod_{i<j} ( \lambda_i -\lambda_j)^2
\exp \left( -\frac12 \sum_i \lambda_i^2 \right)
\prod_{ij} ( 1 - 2 \sqrt{\beta} ( \lambda_i + \lambda_j) )^{-\frac12}
( 1 + 2 \sqrt{\beta} ( \lambda_i + \lambda_j) )^{-\frac12}
\end{equation}
which is 
identical to equation.(\ref{ewot}) when the two measure factors are combined.
This is not so surprising when one realizes that the actions
of equ.(\ref{eXYXY1}) and equ.(\ref{eXYXY2}) are related by
our old friend
the linear transformation $X \to ( X + Y ) / \sqrt{2},
Y \to ( X - Y ) / \sqrt{2}$.

We note in closing that the appropriate linear combination of the first three
terms in equ.(\ref{ePhi4dict}) produces an $X Y X Y$ potential alone,
which was also shown in \cite{14} to display Ising critical behaviour
by carrying out an ingenious {\it non}-linear transformation
on the matrices. Similar models occur in the enumeration of
meanders \cite{15}.

\section{Discussion}

We have considered vertex models on thin $\phi^3$ and $\phi^4$ graphs
and shown that various sorts of critical behaviour -- Ising, Potts,
multi-spin interaction, \ldots  $\;$ may be obtained by restricting
the most general couplings allowed by the arrow reversal symmetry
in the vertex models to various subsets.
In this sense the vertex models discussed here play a similar
role to the eight vertex model on the regular square lattice, in
whose phase diagram lurks the critical behaviour of many other familiar models
\cite{8}.
One important difference from the regular
lattice case is that the 
notion of orientational order for edges round  a vertex 
is lost on random graphs (there is no ``up'' and ``down'').
We showed that writing the vertex model actions in real form 
allowed us to  
recognise transformations from other known models  and deduce
the critical behaviour. The simplicity of the thin graph models
also allowed the explicit solution of  the saddle point equations
in some cases to confirm these deductions, in particular for the various Ising models.
We noted that the vertex models, when written in their equivalent real form,
also admitted a physical interpretation in terms of spin models
on cacti graphs, or loop/line gases on the original random graphs.

Although just as easy to write down actions for fat graphs, solving such
models is a rather more difficult task than on thin graphs. 
We presented some solutions for vertex models on fat graphs  
by showing their 
equivalence to already solved models, mostly by following a
similar tack to the thin graph exposition and rewriting the models
in terms of two Hermitian matrices. In these examples Ising, XY and Ising-like
(for the model of \cite{13}) critical behaviour was observed,
as well as a model that was equivalent to free fermions
(that of equ.(\ref{eXYXY0}), \cite{12a}).
On fat graphs the cyclic order of edges round a vertex may be consistently
defined, so the number of distinguishable terms in higher than $\phi^3$
potentials is greater than in the thin graph case.

As noted in the introduction, the idea of using integrals over
complex variables and saddle point methods to extract
the asymptotic behaviour of directed graph models
for large graphs has been suggested by Whittle \cite{5}.
From a graph theory point of view all of the calculations
in this paper are exercises in counting various sorts
of more or less esoteric directed random graphs.
The use of scalar ``actions'' 
and saddle point methods seems to be
as effective in investigating the statistical mechanics
of vertex models on thin graphs 
(or, if one prefers, the statistical mechanics
of directed graphs) as similar methods
have been for other statistical mechanical, notably spin, 
models living on thin graphs.

DAJ would like to thank Thordur Jonsson for various discussions about
vertex models on planar graphs.

%

%
\clearpage \newpage
\begin{figure}[htb]
\vskip 20.0truecm
\includegraphics{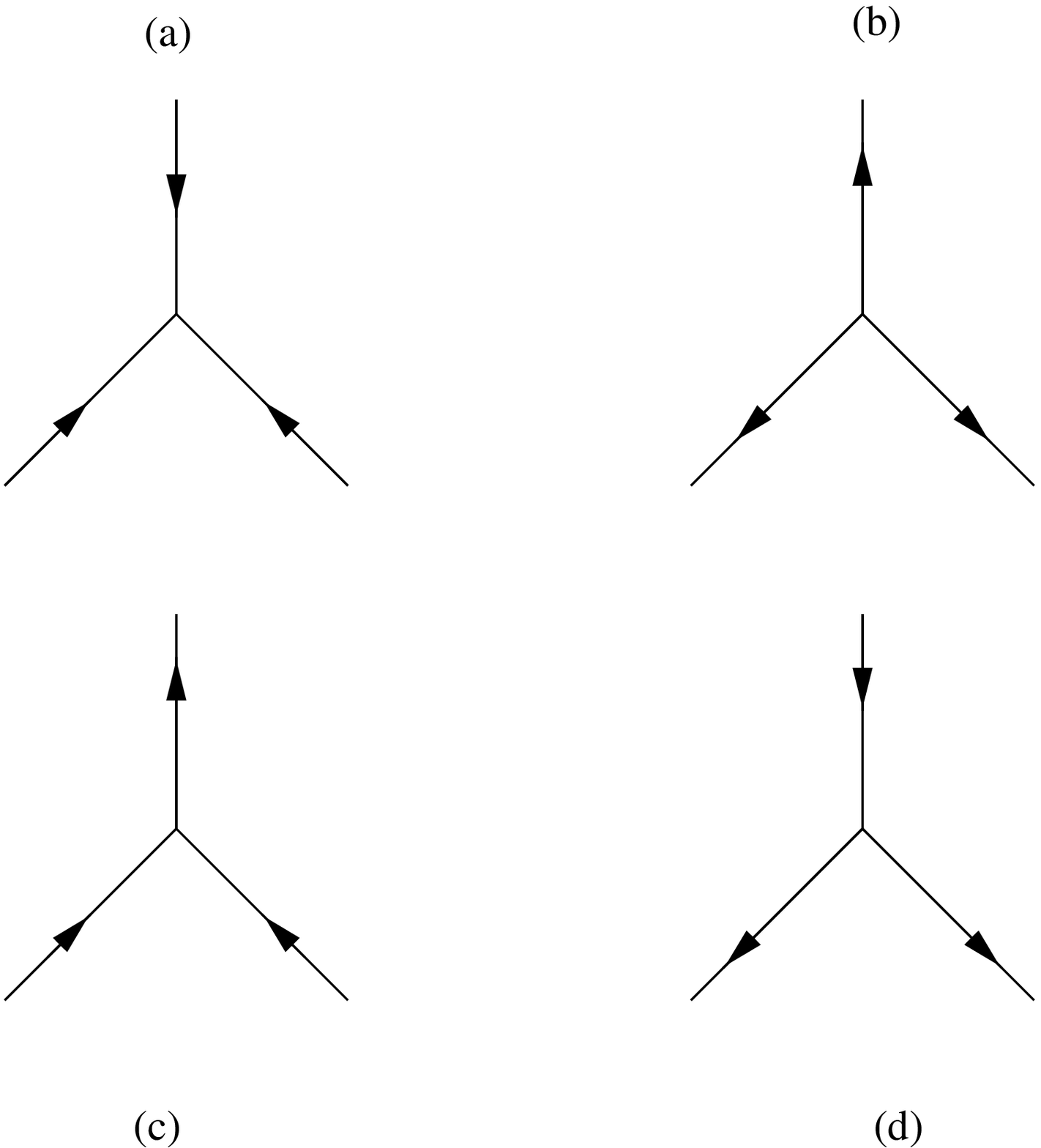}
\caption[]{\label{fig1} The various sorts of $\phi^3$ vertices
which appear in the model:(a)  $(\phi^{\dagger})^3$, (b) $\phi^3$, (c) $(\phi^{\dagger})^2 \phi$, (d) $\phi^2 \phi^{\dagger}$}
\end{figure}
\clearpage \newpage
\begin{figure}[htb]
\vskip 20.0truecm
\includegraphics{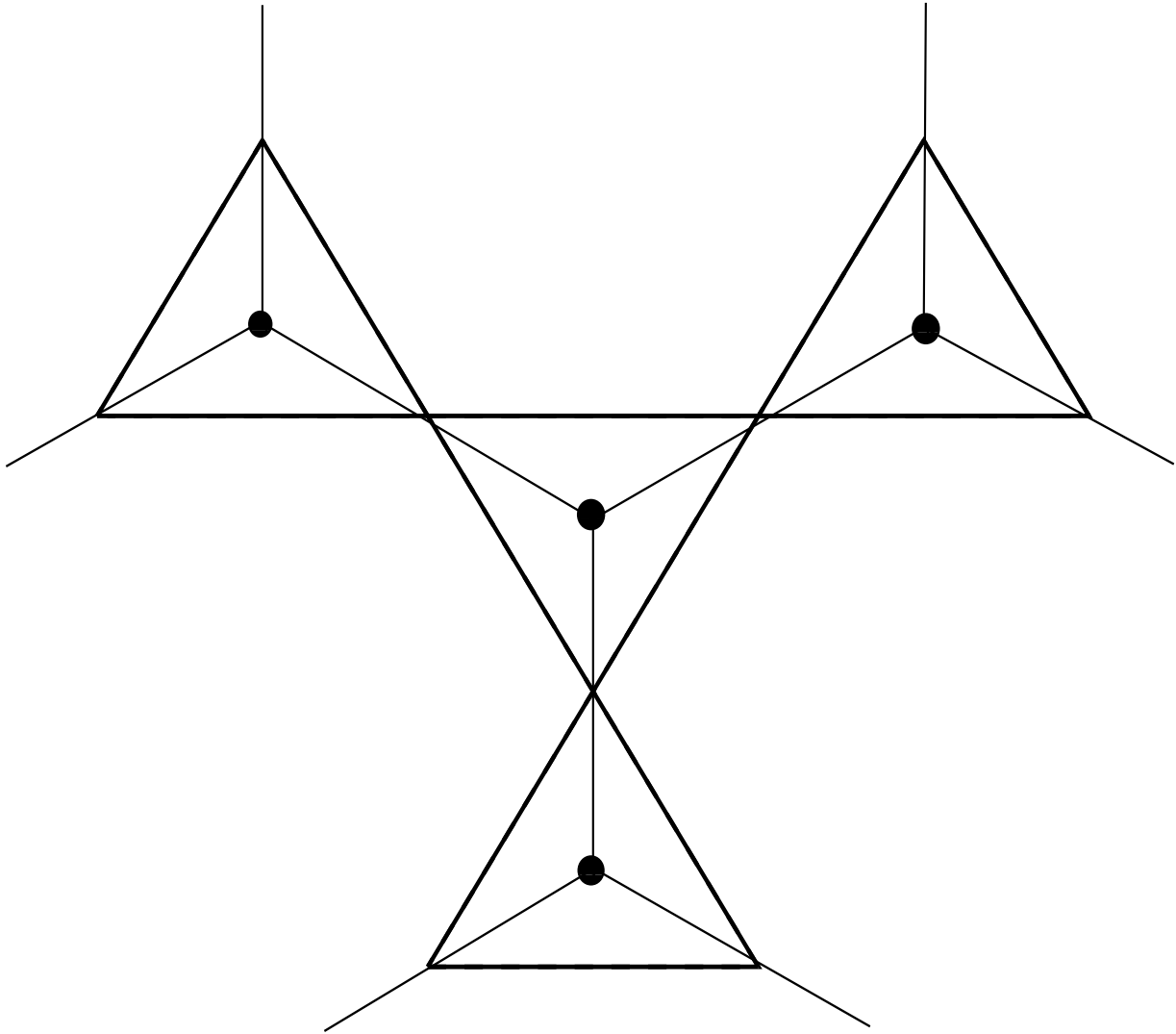}
\caption[]{\label{fig2} A section of a Husimi tree 
composed of triangles. The underlying
$\phi^3$ graph is also shown.}
\end{figure}
\clearpage \newpage
\begin{figure}[htb]
\vskip 20.0truecm
\includegraphics{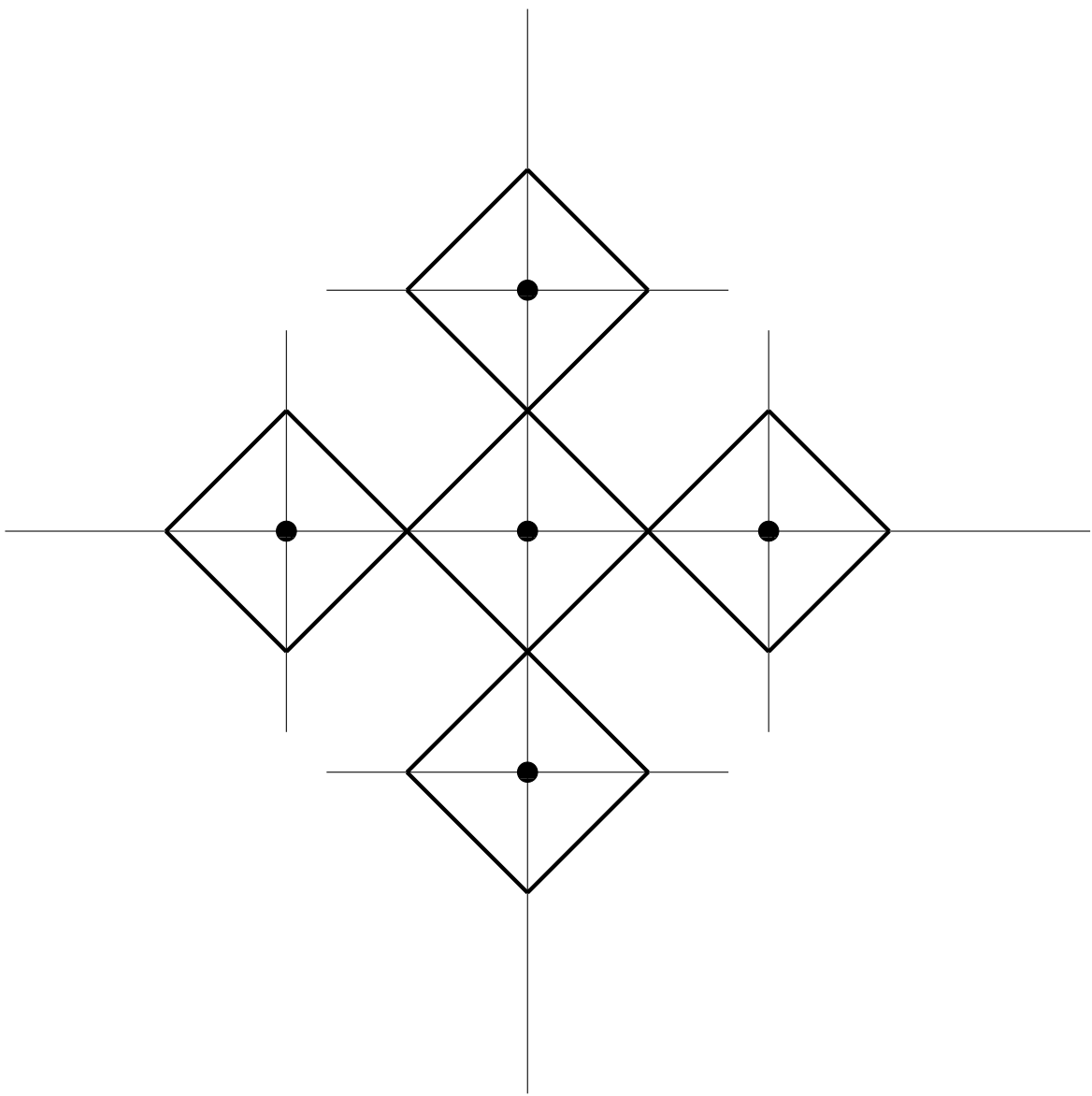}
\caption[]{\label{fig3} A section of a Husimi tree
composed of squares. The underlying
$\phi^4$ graph is also shown.}
\end{figure}
\begin{figure}[htb]
\vskip 20.0truecm
\includegraphics{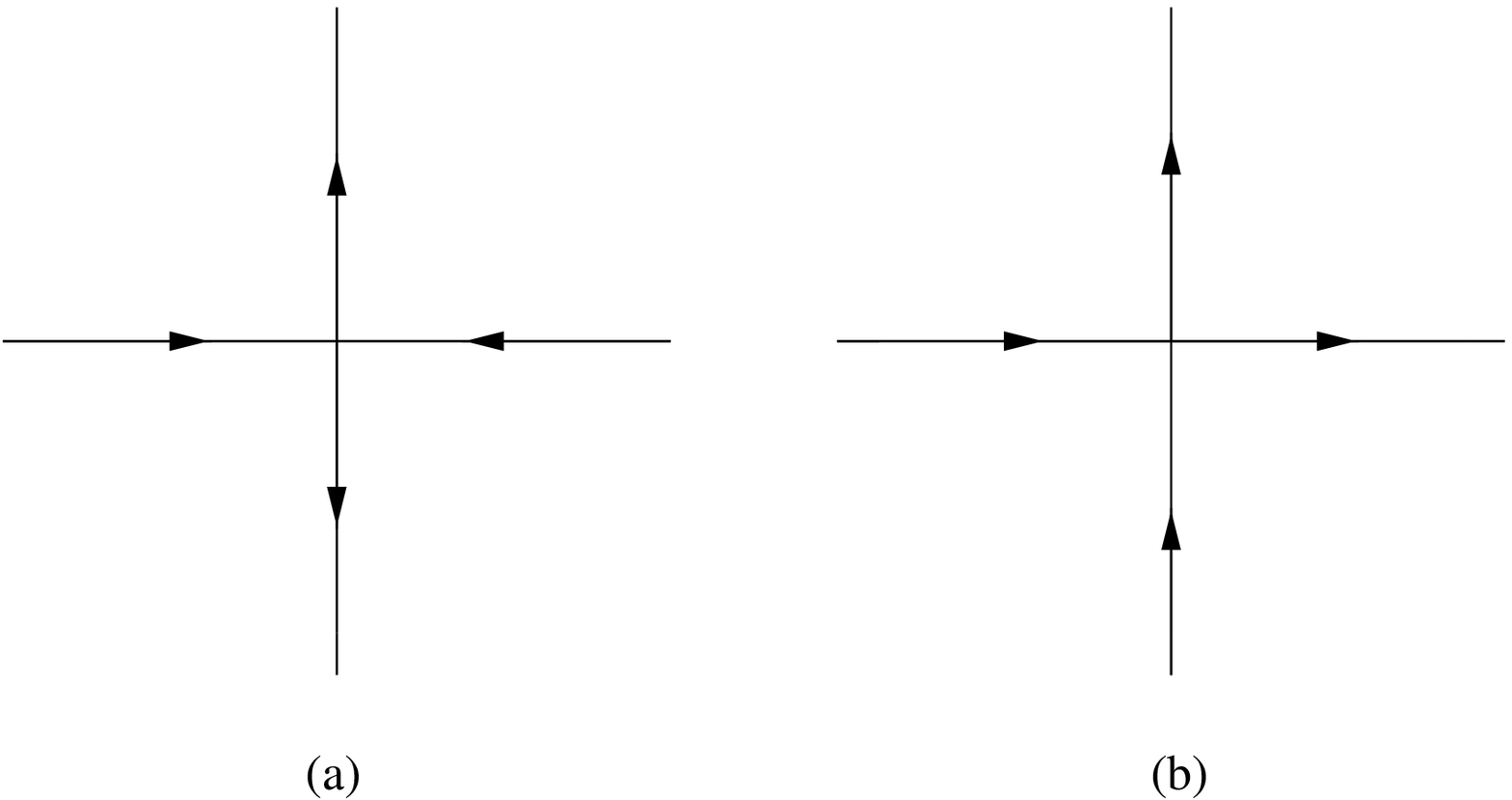}
\caption[]{\label{fig4} Two different cyclic
orientations of a quartic vertex which may be distinguished
in a planar graph: (a) $[ \Phi^{\dagger} \Phi ]^2$, 
(b) $[\Phi^{\dagger}]^2 \Phi^2$ }
\end{figure}

\begin{thebibliography}{99}
\bibitem{1} D. Johnston and P. Plech\' a\v{c}, ``Potts Models on Feynman Diagrams'',
[hep-lat/9704020];\\
D. Johnston and P. Plech\' a\v{c}, ``Percolation on a Feynman Diagram'',
[cond-mat/9705101];\\
D. Johnston and P. Plech\' a\v{c}, ``Why Loops Don't Matter'',
[cond-mat/9705101].
\bibitem{1a}C. Baillie, D. Johnston and J-P. Kownacki, Nucl. Phys. {\bf B432} (1994) 551;\\
            C. Baillie, W. Janke, D. Johnston and P. Plech\' a\v{c}, Nucl. Phys. {\bf B450}
(1995) 730;\\
            C. Baillie and D. Johnston, Nucl. Phys. {\bf B47} (Proc. Suppl.) (1996) 649;\\
            C. Baillie, D. Johnston, E. Marinari and C. Naitza, J. Phys. {\bf A29} (1996) 6683;\\
            C. Baillie, N. Dorey, W. Janke and D. Johnston, Phys. Lett {\bf B369} (1996) 123.
\bibitem{2} C. Bachas, C. de Calan and P. Petropoulos, J. Phys. {\bf A27} 
            (1994) 6121.
\bibitem{3}T. Banks, C. Bender, T-T. Wu, Phys. Rev. {\bf D8} (1973) 3346;\\
          T. Banks, C. Bender, Phys. Rev. {\bf D8} (1973) 3367;\\
               E. Brezin, J. Le Guillou and J. Zinn-Justin,
       Phys. Rev. {\bf D15} (1977) 1544;{\it ibid}
1558;\\
G. Parisi, Phys. Lett. {\bf 66B} (1977) 167;\\
              N. Lipatov, JETP Lett. {\bf 24} (1976) 157;
Sov. Phys. JETP {\bf 44} (1976) 1055; JETP Lett. {\bf 25} (1977) 104;
Sov. Phys. JETP {\bf 45} (1977) 216;\\
               S . Coleman, Phys. Rev. {\bf D15} (1977) 2929;\\
               C. Callan and S. Coleman, Phys. Rev. {\bf D16} (1977)
\bibitem{4}P. Whittle, Adv. Appl. Prob. {\bf 24} (1992) 455;\\
                 in {\it Disorder in Physical Systems}, ed. G.R. Grimmett
and D.Welsh, (1990) 337.
\bibitem{5} P. Whittle, J. Stat. Phys. {\bf 56} (1989) 499.
\bibitem{6}  T. Morris, Nucl. Phys. {\bf B356} (1991) 703.
\bibitem{7}  W. Janke, Phys. Lett. {\bf A143} (1990) 107.
\bibitem{8} R. Baxter, ``Exactly Soluble Models in Statistical Mechanics'', Academic Press, London,
1982.
\bibitem{9} D. Johnston, Phys. Lett. {\bf B314} (1993) 69;\\
            C. Baillie and D. Johnston, Phys. Lett. {\bf B357} (1995) 287;\\
            S. Carroll, M. Ortiz and W. Taylor, Nucl. Phys. {\bf B468} (1996) 620. 
\bibitem{10} X. Wu and F. Wu, J. Phys. {\bf A 22} (1989) L1031;\\
             R. Baxter and F. Wu, Phys. Rev. Lett {\bf 31} (1973) 1294;\\
             C. Thompson, J. Stat. Phys. {\bf 27} (1982) 441; {\it ibid} 457;\\
             J. Monroe, J.Stat. Phys. {\bf 65} (1991) 255; {\it ibid} {\bf 67} (1992) 1185;\\
            P Gujrati, Phys. Rev. Lett. {\bf 74} (1995) 809;\\
            A. Ananikian, S. Dallakian, N. Izmailian and K. Oganessyan, Phys. Lett. {\bf A214} (1996) 205;\\
            A. Alahverdian, N. Ananikian, S. Dallakian, ``Singularities at a Dense Set of Temperature
            in Husimi Tree'', cond-mat/9702106.
\bibitem{11}V.G. Knizhnik, A.M. Polyakov and
            A.B. Zamolodchikov, Mod. Phys. Lett. {\bf A3} (1988) 819.\\
            F. David, Mod. Phys. Lett. {\bf A3} (1988) 1651.\\
            J. Distler and H. Kawai, Nucl. Phys. {\bf B321} (1989) 509.
\bibitem{12} P. Ginsparg, ``Matrix Models and 2D Gravity''
[hep-th/9112013], also in Phys. Rep. {\bf 254} (1995).
\bibitem{12a} S. Dalley, Mod. Phys. Lett {\bf A7} (1992) 1651. 
\bibitem{13} G. Cicuta, L. Molinari and E. Montaldi,
Phys. Lett. {\bf B306} (1993) 245.
\bibitem{14} L. Chekhov and C. Kristjansen, Nucl. Phys. {\bf B479} (1996) 683.
\bibitem{15} O. Golinelli and E. Guitter, Nucl. Phys. {\bf B482} (1996) 497.
          
\end{thebibliography}
\end{document}